\documentclass[fleqn,twoside]{article}
\usepackage{espcrc2}

\usepackage{graphicx}
\usepackage[figuresright]{rotating}

\newcommand{\vek}[1]{\mbox{\bf #1}}

\newcommand{\AmS}{{\protect\the\textfont2
  A\kern-.1667em\lower.5ex\hbox{M}\kern-.125emS}}

\title{Ultrahigh energy cosmic rays from dark matter annihilation}

\author{Rainer Dick\address[UofS]{Department of Physics and Engineering Physics,
        University of Saskatchewan,\\ \hspace{0.5mm} 116 Science Place, Saskatoon, 
        Canada SK S7N 5E2},
        Pasquale Blasi\address[arcetri]{INAF/Osservatorio Astrofisico di Arcetri, 
        Largo E. Fermi 5, 50125 Firenze, Italy}
        and Edward W. Kolb\address[fermilab]{NASA/Fermilab Astrophysics Center,
        Fermi National Accelerator Laboratory,\\ \hspace{0.5mm} Batavia, 
        Illinois 60510-0500}\address[UofC]{Department of Astronomy and Astrophysics,
        Enrico Fermi Institute, The University of Chicago,\\ \hspace{0.5mm}
        Chicago, Illinois 60637-1433}
}
       
\begin{document}

\begin{abstract}
Annihilation of clumped superheavy dark matter provides an interesting explanation for
the origin of ultrahigh energy cosmic rays.

The predicted anisotropy signal provides a unique signature for this scenario.
\vspace{1pc}
\end{abstract}

\maketitle

\section{INTRODUCTION}

Explaining
the origin of cosmic rays at energies above the Greisen--Zatsepin--Kuzmin
bound $E_{GZK}\simeq 4\times 10^{19}\,$eV is an interesting
and puzzling scientific challenge for several reasons:

AGASA observes a flux of order of one cosmic ray
event per $100\,\mathrm{km}^2$ per year at energies
above $10^{20}\,$eV \cite{AGASA},
thus confirming the flux seen by previous smaller ground based
arrays on the northern hemisphere\footnote{But see \cite{watson2} for a
recent discussion of the pertinent problem to reconstruct primary cosmic
ray energies from air showers.
A new analysis presented there lowered the estimate on the energy of the highest
energy event observed with the Haverah Park Array from above $10^{20}\,$eV
to $8.3\times 10^{19}\,$eV and also indicates a smaller flux at the high
energy end of the spectrum. Of course, the very existence of UHECRs is not
questioned by these findings.}, see e.g. \cite{NW}.

On the other hand,
energy loss of electromagnetically interacting particles due to scattering
with the CMB background indicates that these ultrahigh energy
cosmic rays (UHECRs) should originate within a radius
of order $100\,$Mpc \cite{GZK,new,bbo}, and sufficiently powerful
astrophysical acceleration mechanisms are not easy to find.

This has motivated proposals that superheavy dark matter may be the
origin of ultrahigh energy cosmic rays:

This includes in particular
Hill's proposal of {\em monopolonium decay} \cite{hill}, the 
{\em decay of superheavy dark matter relics} proposed by 
Berezinsky {\it et al.} \cite{BKV} and by Kuzmin and Rubakov \cite{KR},
and most recently our proposal of {\em collisional annihilation
of superheavy dark matter} \cite{bdk1,bdk2}. 
While 
decay scenarios for UHECRs should imply domination of the galactic
center in the distribution of arrival
directions, the proposal of collisional annihilation
predicts that with sufficient statistics
from the upcoming Auger observatory
a clumpy structure should emerge in the galactic source distribution:
Unitarity bounds on reaction cross sections imply that a smooth
superheavy dark matter distribution in the galactic halo would not yield
the observed flux through collisional annihilation, but the overdensity
in dark matter clumps in the halo may well account for the 
observed flux \cite{bdk1}.

The common virtues of these top-down scenarios are:\\
$\bullet$ a natural explanation of the seemingly random distribution
of arrival directions;\\
$\bullet$ avoidance of collisional damping and
the GZK bound due to origination in the galactic
halo;\\
$\bullet$ no need for an extremely powerful astrophysical acceleration
mechanism.

Very high-energetic neutrinos both from superheavy dark matter decay
and from annihilation have very recently been discussed in \cite{francis},
and some recent overviews of top-down scenarios for UHE cosmic rays
can be found in \cite{gabor,pasquale,Sark3}.

Superheavy dark matter particles cannot arise as a consequence of standard
thermal freeze out, but it has been pointed out that dark
particles might have never experienced local chemical equilibrium during the
evolution of the universe, and that their mass may be in the range $10^{12}$ to
$10^{19}$ GeV, much larger than the mass of thermal {\sc wimps}
\cite{ckr1,ckr2,ckr3,ckr4}. Since these {\sc wimps} would be much more massive than
thermal {\sc wimps}, such superheavy dark particles have been called 
{\sc wimpzillas} \cite{ckr4}.

{\sc wimpzillas} may be created during 
bubble collisions if inflation is completed
through a first-order phase transition \cite{barrow,mr}; at the preheating
stage after the end of inflation with masses easily up to the GUT
scale \cite{Kolb:1998he} or even up to the Planck scale
\cite{CKRT,Giudice:1999fb}; or during the reheating stage after 
inflation \cite{ckr3}.
{\sc wimpzillas} may also be generated in the transition between an inflationary
and a matter-dominated (or radiation-dominated) universe due to the
``nonadiabatic'' expansion of the background spacetime acting on the vacuum
quantum fluctuations \cite{ckr1,kuzmin,cckr}. The distinguishing feature of 
this mechanism is the 
capability of generating particles with mass of the order of the inflaton mass
even when the particles only
interact extremely weakly (or not at all) with other particles.

The mass $m$ of unstable dark matter which decays through gravitational couplings
is constrained by the fact that the lifetime is only of 
order $\tau\le m_{Planck}^2/m^3$ \cite{rd1,rd2}, and therefore unstable heavy
relics must decay through strongly suppressed channels which effectively
correspond to sub-gravitational couplings. This provided one motivation for
the collisional annihilation proposal in \cite{bdk1}, since the need for
2-particle collisions implies a larger lifetime of annihilating 
{\sc wimpzillas} as opposed to decaying superheavy particles.

\section{THE FLUX FROM ANNIHILATION IN DARK MATTER CLUMPS}

The spectral fluxes from decay or annihilation of {\sc wimpzillas} 
of mass $M_X$ and particle density $n_X(\vek{r})$ are
\begin{equation}\label{eq:fluxdec}
\mathcal{F}_d(E)=\frac{d\mathcal{N}(E,M_X/2)}{dE}\frac{1}{\tau_d}
\int\! d^3\vek{r}\,\frac{n_X(\vek{r})}{2\pi|\vek{r}_{\odot}-\vek{r}|^2}
\end{equation}
and
\begin{equation}\label{eq:fluxann}
\mathcal{F}(E)=\frac{d\mathcal{N}(E,M_X)}{dE}\langle\sigma_A v\rangle
\int\! d^3\vek{r}\,\frac{n_X(\vek{r})^2}{2\pi|\vek{r}_{\odot}-\vek{r}|^2},
\end{equation}
respectively.
The shape of the spectrum is determined by the
 function $d\mathcal{N}(E,E_\textrm{\small jet})/dE$, which
gives the number of sufficiently long-lived particles per energy interval
emerging from a primary jet of energy $E_\textrm{\small jet}$.
Birkel and Sarkar and Berezinsky and Kachelriess
have used Monte Carlo simulations
to calculate the spectrum from decay of superheavy relics
\cite{BSark,BK}. Fodor and Katz as well as Sarkar and Toldr\`{a}
employed numerical integrations
of the DGLAP evolution equations \cite{FK,ST}.
The resulting spectrum in the interesting range
between $10^{19}\,$eV and $10^{20}\,$eV is similar to the spectrum
from the old modified leading
log approximation (MLLA) \cite{mlla}, which was
employed by Berezinsky et al.\ in their initial proposal of UHE cosmic rays
from decay of superheavy dark matter decay \cite{BKV}.

\begin{figure}[htb]
\begin{center}
\resizebox{0.45\textwidth}{!}{ \includegraphics{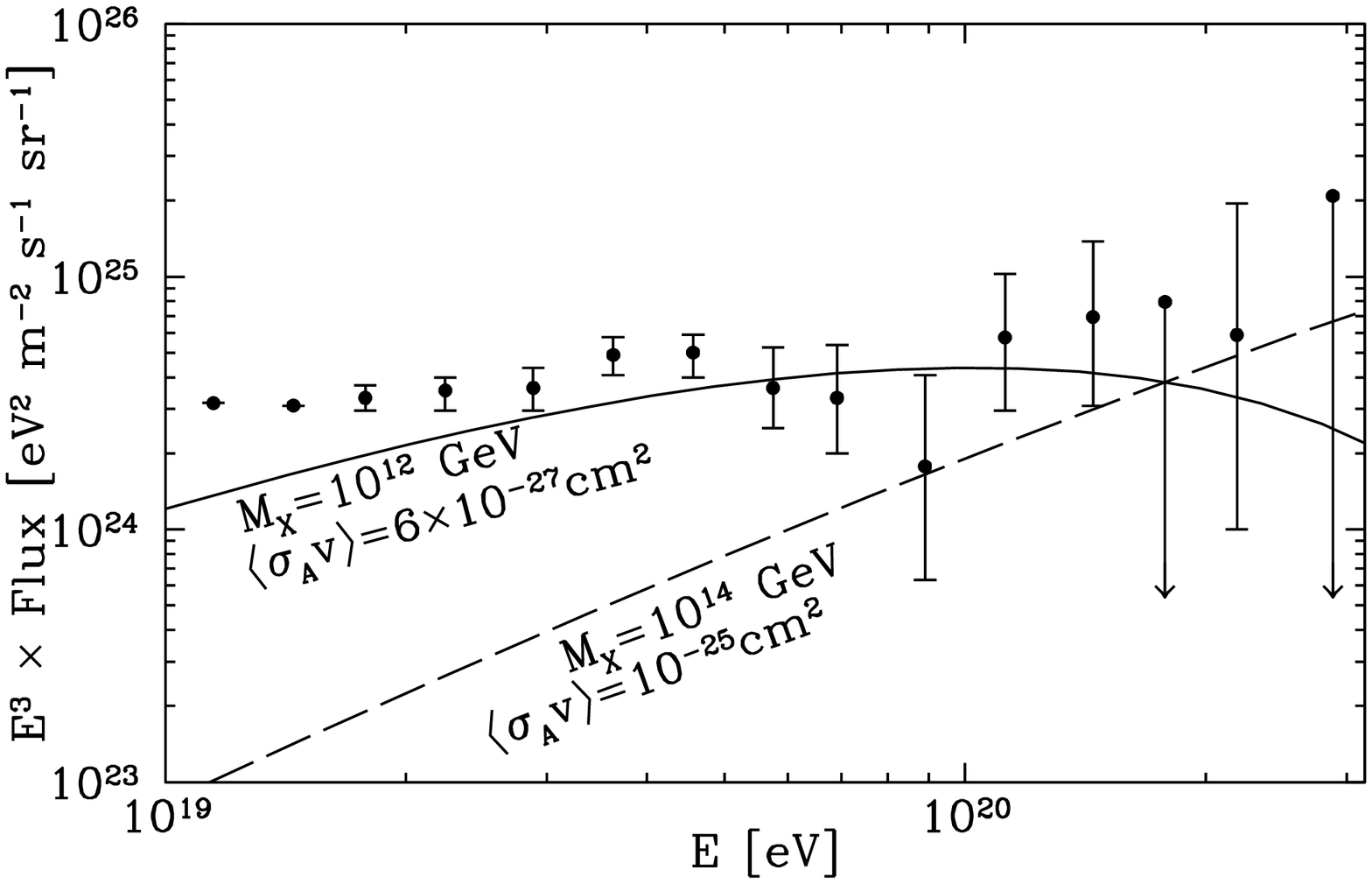} }
\resizebox{0.45\textwidth}{!}{ \includegraphics{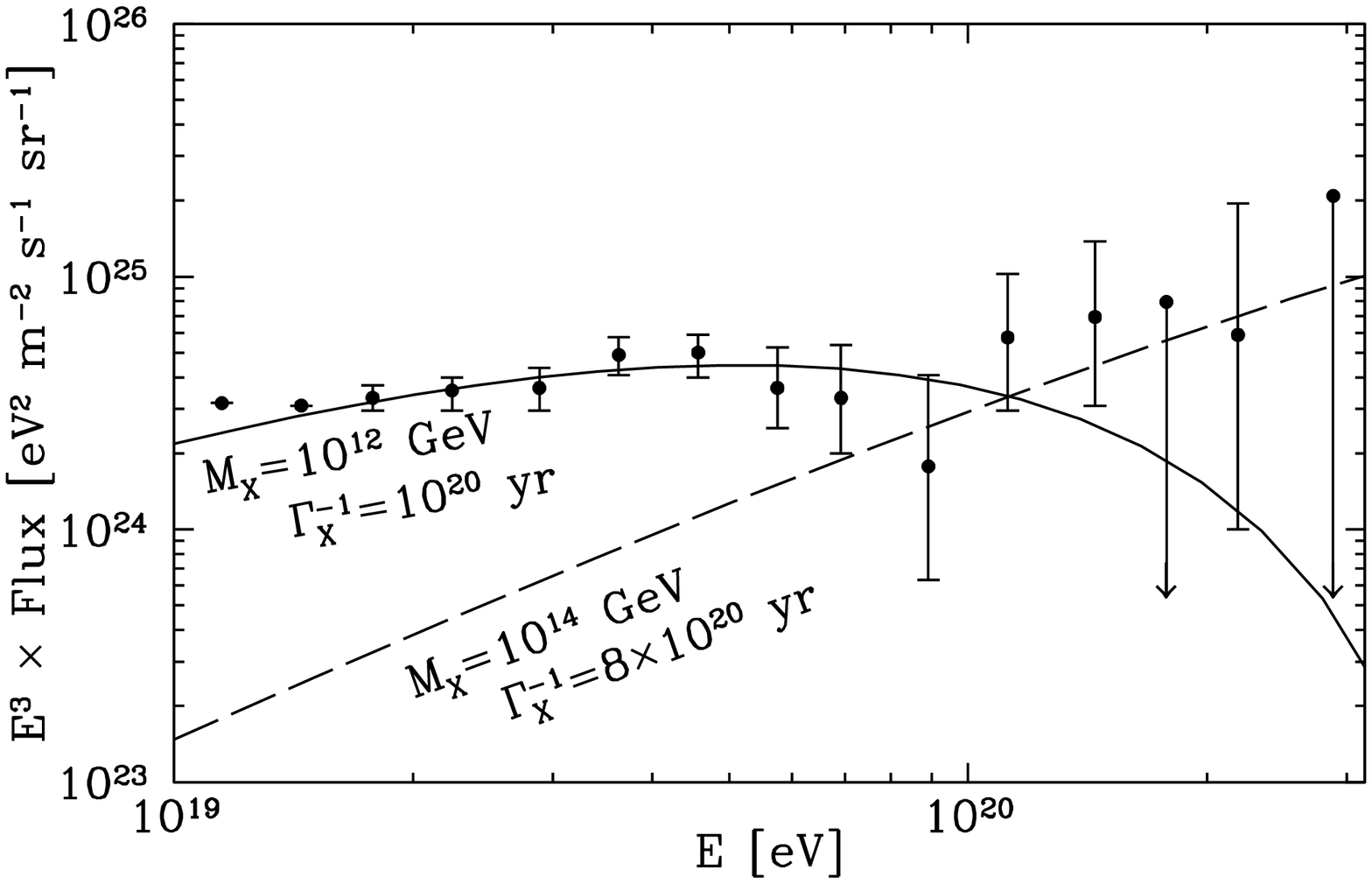} }
\end{center}
\vspace*{-10mm}
\caption{
UHE cosmic ray spectra from superheavy particle annihilation (upper
panel) or decay (lower panel). 
}
\label{spectrum}     
\end{figure}

We had also used the MLLA limiting spectrum in our proposal of 
{\sc wimpzilla} annihilation \cite{bdk1} (Fig.\ \ref{spectrum}).

The annihilation cross sections from the normalization
in Fig.\ \ref{spectrum} were calculated for
a smooth NFW \cite{NFW} background halo. 
These cross sections would be in conflict with unitarity bounds 
on reaction cross sections for particles of mass $M_X\ge 10^{12}\,$GeV and
therefore the flux from {\sc wimpzilla} annihilation in the smooth halo
component will (most likely) be negligible\footnote{There is also
the possibility that {\sc wimpzillas} form meta-stable
bound states, similar to monopolonium. Then the contribution from the smooth
halo component could be important and would resemble
the anisotropy signature expected from monopolonium or superheavy dark matter
decay.}.
However, a fraction around a few percent of the dark matter
in the halo is expected to exist in dense isothermal dark matter 
sub-clumps\footnote{See e.g.\ \cite{moore} for evidence for 
sub-clumps from $N$-body simulations, and \cite{DK} for evidence from
gravitational lensing.},
and due to large overdensities in the cores {\sc wimpzilla} annihilation
in these sub-clumps can comfortably generate the observed UHE flux
with annihilation
cross sections well below the unitarity bounds \cite{bdk1}.
This does not change the good fit of the annihilation model
to the spectrum, which is a concequence 
of $d\mathcal{N}(E,M_X)/dE$, since the difference between events originating
in the smooth and clumpy components affects only the overall factor
$\int\! d^3\vek{r}\, n_X(\vek{r})^2|\vek{r}_{\odot}-\vek{r}|^{-2}$ in 
(\ref{eq:fluxann}).


Due to the negligibility of a possible contribution from the
smooth halo component a clumpy source structure will be an unmistakable
signature of collisional
{\sc wimpzilla} annihilation as a source of UHE cosmic rays. In particular,
nearby sub-clumps will appear as hot spots in the sky (see Fig.\ 4
in Ref.\ \cite{bdk1}).

To calculate the expected overall anisotropy (after averaging over
the clumps in the case of annihilation) one can integrate the local decay
or annihilation rates entering in (\ref{eq:fluxdec}) and 
(\ref{eq:fluxann}) along lines through the galactic halo, taking into
account the galactic halo density and the expected
sub-halo distribution, respectively. 

The left panel in Fig.\ 2 shows the result for annihilation in a relatively
broad subhalo distribution with the sub-clump density dropping off at 
a characteristic radius $R_c^{cl}=20\,$kpc (see
\cite{bdk1} for a discussion of the expected
sub-halo distribution).

\vspace*{15mm}
\begin{minipage}[t]{0.25\textwidth}
\begin{center}
\scalebox{0.2}{\includegraphics{aniplot2.eps}}
\end{center}
\end{minipage}
\hfill
\begin{minipage}[t]{0.25\textwidth}
\begin{center}
\scalebox{0.2}{\includegraphics{aniplot3.eps}}
\end{center}
\end{minipage}

\noindent
Figure 2.\ The left panel shows the envelope of the 
UHE flux from annihilation
in a galactic 
subhalo distribution as a function of the angle towards
the galactic center, whereas the right panel shows the UHE flux from decay
in a 
galactic NFW profile.

\section{CONCLUSION}

All top-down scenarios for generation of UHECRs from superheavy dark matter
predict a large photonic component in cosmic ray primaries above
$E_{GZK}\simeq 4\times 10^{19}\,$eV, and a
more or less dominant role of the galactic center:

Both annihilation in bound states (monopolonium or {\sc "wimpzillonium"} decay)
and decay of superheavy relics imply that 
the dominant smooth dark matter halo distribution
would be the primary source of UHECRs, whence the distribution of UHECR
arrival directions should track the overall dark matter halo.

Collisional annihilation, on the other hand, would almost exclusively appear
in the cores of local dark matter overdensities, and therefore would
primarily track local overdensities. Here some domination 
of arrival directions pointing back towards the galactic
center is expected because the sub-clump population is expected to drop
off with distance from the galactic center.
A nearby clump out of the line of sight to the galactic center
could dominate the local signal and appear as a hot spot.

The  southern site of the Auger observatory sees
the galactic center, and therefore the prediction of a clumpy
source distribution with some accumulation towards the galactic
center should be testable after about three years of full
operation of the southern site, which should yield of the order
of 100 UHECR events with energies at and beyond $10^{20}\,$eV.

\end{document}